# Over-limiting Current and Control of Dendritic Growth by Surface Conduction in Nanopores

*Ji-Hyung Han[1], Edwin Khoo[1], Peng Bai[1], and Martin Z. Bazant[1,2]\**

Departments of Chemical Engineering[1] and Mathematics[2]

Massachusetts Institute of Technology, Cambridge, MA 02139, USA

SUMMARY

Understanding over-limiting current (faster than diffusion) is a long-standing challenge in electrochemistry with applications in desalination and energy storage. Known mechanisms involve either chemical or hydrodynamic instabilities in unconfined electrolytes. Here, it is shown that over-limiting current can be sustained by surface conduction in nanopores, without any such instabilities, and used to control dendritic growth during electrodeposition. Copper electrodeposits are grown in anodized aluminum oxide membranes with polyelectrolyte coatings to modify the surface charge. At low currents, uniform electroplating occurs, unaffected by surface modification due to thin electric double layers, but the morphology changes dramatically above the limiting current. With negative surface charge, growth is enhanced along the nanopore surfaces, forming surface dendrites and nanotubes behind a deionization shock. With positive surface charge, dendrites avoid the surfaces and are either guided along the nanopore centers or blocked from penetrating the membrane.

Many industrial processes rely on electrodeposition to make smooth metal coatings, but uniform electroplating is often unstable to the growth of finger-like dendrites. For over three decades, dendritic copper electrodeposition has been studied as an example of diffusion-limited fractal growth[1,2], although it has become clear that electric fields and fluid flows also play important roles[3-5]. Suppressing dendrites is a critical challenge for lithium-ion[6,7] and lithium-air[8,9] batteries, in order to prevent capacity losses and catastrophic short circuits during recharging, which can be mitigated by electrolyte design[10,11] or nanostructured separators[12-14] and surface layers[15]. On the other hand, dendritic short circuits can also be exploited for sensing and information storage[16-19], if well controlled at the nanoscale. Dynamical control of electrodeposition is also critical for the fabrication of nanostructures[20-25], nano-electronics[23,26], 3D integrated circuits[27], and 3D batteries[28-30].

Dendritic growth allows an electrode to overcome diffusion limitation at high currents by focusing the ionic flux on rapidly advancing dendrite tips[1-4, 31-33]. In an unsupported binary electrolyte, driving current into a cation-selective surface, such as an electrode or membrane, depletes the salt concentration, as cations are removed and anions repelled to maintain electroneutrality. Classical theories of ion concentration polarization predict a diffusion-limited current[34], but "over-limiting current" (OLC) faster than diffusion has long been observed in electrodialysis[35-38] and nanofluidics[39] and investigated for desalination[35, 38] and fuel cells[40].

Theoretical mechanisms for OLC involve either electrochemical reactions or transport processes other than electro-diffusion that replenish the salt concentration[38]. Electrochemical mechanisms include water splitting[38] and current-induced membrane discharge[41]. A fundamental hydrodynamic mechanism observed in electrodialysis[38] and nanofluidics[42] is the electro-osmotic instability (EOI) of Rubinstein and Zaltzman[38, 43]. EOI results from second-kind electro-osmotic slip in the electric double layer (EDL) on the ion-selective surface[43], leading to convection[39] and chaotic flows. EOI has been observed near a membrane with tracer particles[36] and near a single nanoslot with fluorescent molecules[37] and is affected by inhomogeneous conductivity. In microchannels, multiple vortices and concentration plateaus have been observed in the ion depletion region[44], which do not occur in an unconfined electrolyte, according to theory[38, 43] and experiments[38].

Under confinement in a channel or pore with charged surfaces, Dydek et al.[45] have predicted transitions from EOI to two new mechanisms for OLC, electro-osmotic flow (EOF) and surface conduction (SC), as the channel thickness is decreased. The EOF mechanism, first suggested by Yaroshchuk et al.[46], is based on surface convection[45] that leads to "wall fingers" of salty fluid reaching the membrane without diffusive mixing[47]. The first experimental evidence for the EOF mechanism was recently reported by Deng et al.[35] using a silica glass frit, where surface convection leads to "eddy fingers" in the porous network. A hallmark of the EOF mechanism is the persistence of OLC if the sign of the surface charge is flipped, thereby reversing the EOF vortices[35]. According to the theory[45], EOF plays a larger role than EOI in microchannels[39, 44], but SC should dominate in nanochannels[45] where transverse diffusion suppresses surface convection. Dydek et al.[45] noted that this transition is suggested by microfluidic particle-tracking experiments[39], but the SC mechanism remains to be confirmed experimentally.

Without probing the dynamics at the pore scale, Archer and co-workers have recently shown that charged nanoporous polymer/ceramic separators can help to stabilize electrodeposition in rechargeable lithium metal batteries[12-14]. The introduction of ceramic particles or porous solids with tethered ionic-liquid anions was shown to improve cycle life by reducing dendritic growth. Besides mechanical blocking of dendrites, it was conjectured that dendritic instability is suppressed by the reduction of space charge at the metal/solution interface[48, 49]. This hypothesis refers to yet another mechanism for OLC, the formation of an extended non-equilibrium double layer, which could theoretically occur at a membrane[50] or electrode[51], but only in the absence of EOI[43]. Indeed, the electro-convection observed at dendrite tips is inconsistent with extended space charge[4] and is likely attributable to EOI since the linear growth instability can be explained by electro-neutral diffusion[52, 53]. Recently, Tikekar et al.[54] added uniform background charge to the electro-neutral linear stability analysis and found that negative charge enhances the stability of cation electrodeposition. This different explanation of Archer's results is consistent with the predicted stability[55] of the deionization shock (or "diffusive wave"[32, 33]) that would precede the growth in a negatively charged porous medium[55-57] or microchannel[58-60]. The precise role of surface charge on electrodeposition in porous media, however, is neglected by existing models[61,][62] and remains to be established experimentally.

In this letter, we provide electrochemical and visual evidence that SC is the dominant mechanism for OLC in nanopores and investigate its effects on electrodeposition. Our model system is a commercial anodized aluminum oxide (AAO) membrane with nano-sized straight parallel pores (300-400 nm in diameter, 60 μm in thickness, 0.25-0.50 in porosity) whose surface charge is modified with multiple layers of charged polyelectrolytes and used as a template for copper electrodeposition from copper sulfate ($CuSO_4$) solutions. Shin et al. have recently demonstrated diffusion-limited nanowire growth in the same system[25], but without varying voltage or current, and, as in all prior work, the template surface charge was neither controlled nor thought to play any role.

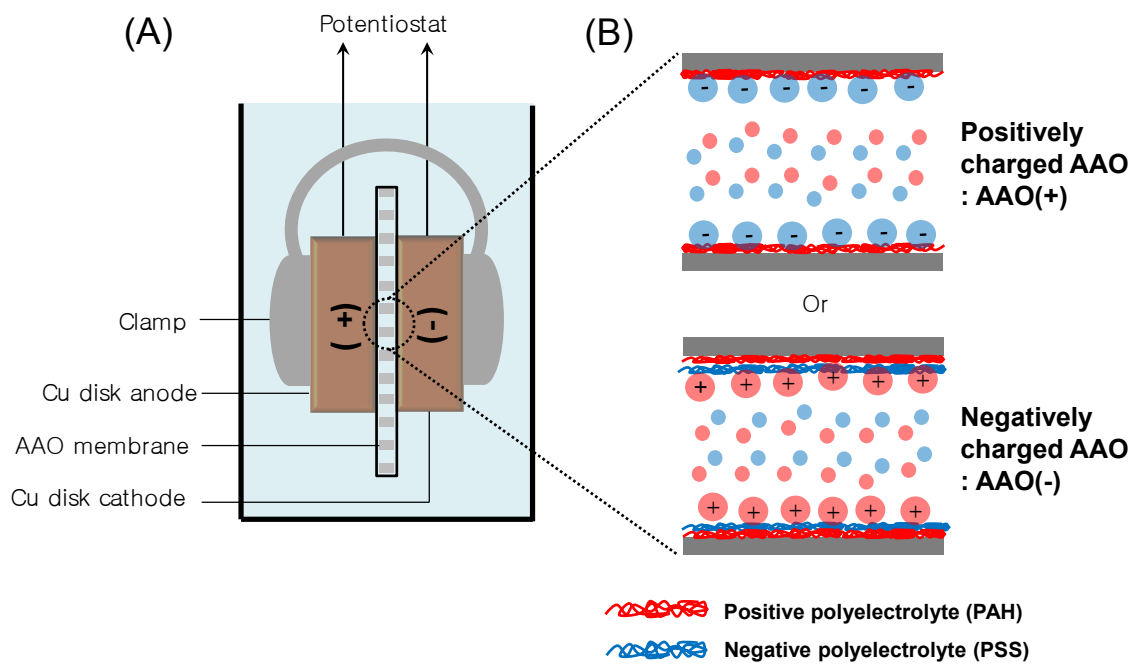

**Figure 1.** (A) Cell configuration in $CuSO_4$ solution: Cu cathode / polyelectrolyte-coated AAO / Cu anode. (B) Nanopore EDL structure. The EDL counter-ions contributing to surface conductivity are displayed as larger circles than the bulk ions.

In our experiments, the AAO membrane is clamped between two copper disk electrodes under constant pressure, as shown in Figure 1A. Electrochemical transient signals are measured in $CuSO_4$ solutions of varying salt concentration, where the dominant Faradaic reactions are copper electrodeposition at the cathode and copper dissolution at the anode. Although the more common method of fabricating the cathodes is to sputter gold or copper onto one side of the AAO membrane, the clamping procedure we use ensures that the shape of the sputtered metal is not a confounding variable that affects the current and the morphology of the electrodeposits[20]. We confirmed that there are no cracks on the AAO membrane when the cell is disassembled after electrochemical measurements. In order to prevent the evaporation of the binary electrolyte solution inside the AAO membrane, the electrochemical cell is immersed in a beaker containing the same electrolyte.

Before assembling the cell, charged polyelectrolyte multilayers are deposited on the side walls of the AAO membrane using the layer-by-layer method[63], based on electrostatic forces between oppositely charged species. Overcompensation of the outer layer causes a dramatic change in the surface potential. This coating method is very versatile and can tune the surface charge of most substrates, including AAO[64]. Positive polyelectrolytes (poly(allylamine hydrochloride), PAH) are directly deposited on the air plasma-treated AAO to confer a positive surface charge, AAO(+). Negatively charged AAO(-) is obtained by depositing negative polyelectrolytes (poly(styrenesulfonate), PSS) on the PAH-coated AAO. Due to the high surface charge density of the layer-by-layer thin films, we expect excess sulfate anions and copper(II) cations to dominate the EDL of the AAO(+) and AAO(-), respectively (Figure 1B).

Across all the experimental conditions of surface charge and salt concentration, the Debye screening length (<10 nm) is small compared to the pore size, but surprisingly such a thin EDL can still dominate ion transport at high voltage. The charged AAO acts as a "leaky membrane"[57], whose neutral salt can be fully depleted near the cathode, leaving SC to support OLC[45] and deionization shocks[35, 55, 56, 58] in AAO(-) or block transport in AAO(+). This interplay between bulk and surface conduction is very different from polyelectrolyte multilayer-coated nanopores with strong EDL overlap, where current rectification is observed[65].

Figure 2A shows experimental current-voltage curves (solid lines) of AAO(+,-) in 10 mM $CuSO_4$ for a linear voltage sweep of 1 mV/s, close to steady state. At low voltage below -0.1 V, the two curves overlap, indicating that the surface charge plays no role, consistent with the classical theory. Unlike ion-exchange membranes[36, 39, 45], a positive curvature is also observed at low voltage, due to the activated kinetics of charge transfer and nucleation. As expected, the onset potential of Cu reduction does not depend on the AAO surface charge.

As the applied potential is increased, dramatic differences in current are observed between AAO(+) and AAO(-). The current in AAO(+) reaches -2.5 mA around -0.2 V and slowly decreases to a limiting current around -2.0 mA, but AAO(-) shows a dramatic linear increase of OLC. The EOI mechanism can be ruled out since it is suppressed in nanopores and insensitive to their surface charge, but EOF could play a role. Since EOF vortices arise regardless of the sign of the surface charge, some OLC can be observed even when the surface charge is reversed, as recently demonstrated for glass frits with micron-scale pores[35]. The lack of any OLC for AAO(+) thus rules out the EOF mechanism.

Instead, the data are consistent with the SC mechanism, as predicted theoretically[45]. The physical picture is sketched in Figure 2B. For AAO(-), SC provides a short-circuit path for $Cu^{2+}$ counter-ions to circumvent the depleted region and reach the cathode by electro-migration in the large local electric field, as $SO_4^{2-}$ co-ions are pushed toward the anode. The EDL thus acts like a shunt resistor around a diode in reverse bias[45]. For AAO(+), the active $Cu^{2+}$ ions are the co-ions repelled from the EDL, while the $SO_4^{2-}$ counterions migrate away from the cathode and further block $Cu^{2+}$ in the diffusion layer in order to maintain neutrality, thus reducing the limiting current.

In order to predict the OLC due to SC, the system can be modeled as a one-dimensional "leaky membrane" governed by Nernst-Planck equations for dilute, electro-neutral ion transport in a constant background charge[35, 45, 55, 57]. The current-carrying cupric ion has valence $z = 2$ and diffusivity $D_0 = 7.14 \times 10^{-10}$ m$^2$/s[66]. Estimates of the negative and positive surface charge densities, -0.75 e/nm$^2$ and 0.375 e/nm$^2$ respectively, are taken from the literature on PAH/PSS polyelectrolyte multilayers[65]. Butler-Volmer kinetics are assumed for copper electrodeposition from copper sulfate solutions with parameters averaged from literature values[67-69] (exchange current density $I_0$ = 2.95 mA/cm$^2$ at 75 mM and symmetry factor $\alpha$ = 0.75), The electrode surfaces move at the same constant velocity, set by the applied current and copper's density, neglecting the porosity of cathode growth at high voltage (described below).

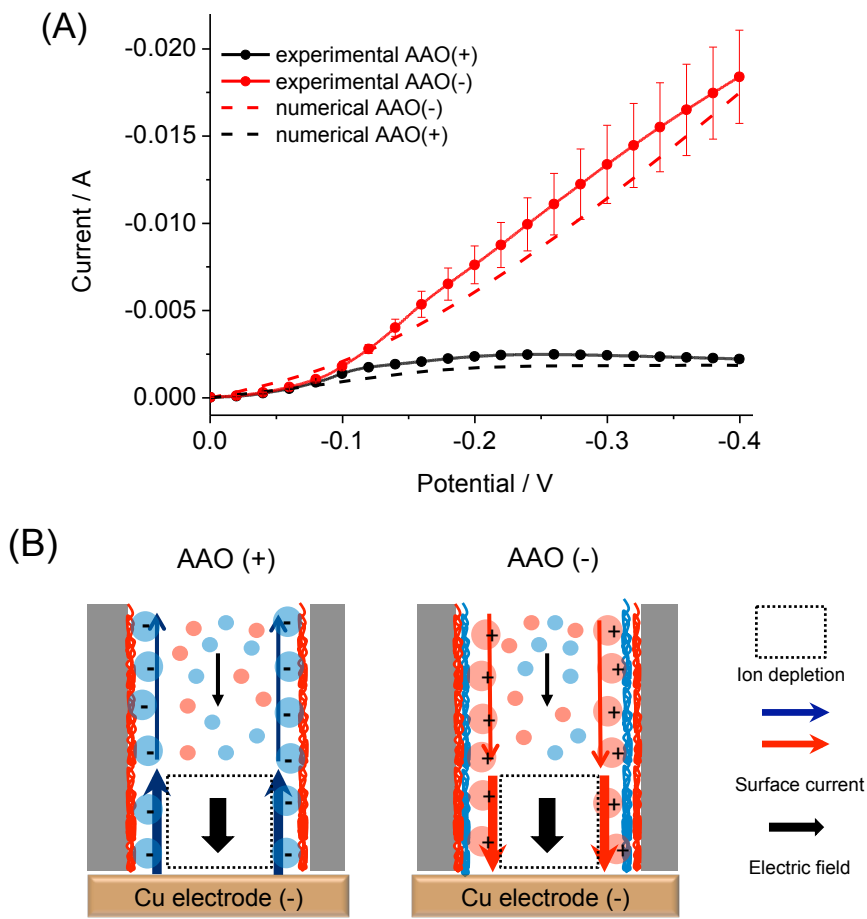

**Figure 2.** (A) Experimental (solid line) and numerical current (dash line) versus voltage data for positively (+) and negatively (-) charged AAO membranes in 10 mM CuSO$_4$ at a scan rate of 1 mV/s. (B) Physical picture of surface conduction effects at high voltage, driven by the large electric field in the depleted region. In AAO(+), the SO$_4^{2-}$ counter-ions (blue) migrate toward the anode, reducing the net flux of Cu$^{2+}$ in order to maintain neutrality. In AAO(-), the active Cu$^{2+}$ counter-ions (red) circumvent the depleted region by SC and contribute to OLC.

This simple model is quantitatively consistent with the data, as shown by numerical solutions in Figure 2A, without adjusting any parameters. To our knowledge, this is the first experimental evidence for OLC due to SC, further corroborated below by impedance spectroscopy and electrodeposit imaging. Analytical predictions can also be derived to better understand scaling relationships. Neglecting SC, the diffusion-limited current is

$$I_{\text{lim}} = \frac{4ze\varepsilon_p D_0 c_0 A}{\tau L} \quad (1)$$

which is twice as large as for a reservoir in place of the anode[35, 45, 55, 57]. For a leaky membrane of length thickness $L = 60$ μm, electrode area $A$, porosity $\varepsilon_p = 0.375$ and tortuosity $\tau = 1$ (straight parallel pores) filled with an electrolyte of mean concentration $c_0 = 75$mM, Equation (1) predicts $I_{\text{lim}} = 3.90$mA, which is close to what is observed experimentally. This supports recent scaling evidence for diffusion-limited dynamics in this system[25], as well as the hypothesis that larger limiting currents observed in random porous media reflect eddy dispersion[34], which cannot occur in the straight, non-intersecting pores of AAO.

The experiments and simulations both show a constant over-limiting conductance $\sigma_{\text{OLC}}$ at high currents defined by $I \sim \sigma_{OLC} * V$, consistent with the SC theory. In this regime, Butler-Volmer kinetics are fast, and the model can be solved analytically. The over-limiting conductance due to SC turns out to be the same as if the anode were replaced by a reservoir[35, 45, 55, 57],

$$\sigma_{\text{OLC}} = \frac{zeD_0 A \epsilon_p \sigma_s}{\tau L k_B T h_p} \quad (2)$$

where $\sigma_s$ is the surface charge density and $h_p$ is the effective pore size, equal to half the pore radius for straight parallel pores. Equation (2) predicts an overlimiting conductance of 0.05395 $\Omega^{-1}$, which is close to the experimental and numerical values, 0.05640 $\Omega^{-1}$ and 0.05329 $\Omega^{-1}$ respectively, further supporting the theory of OLC by SC.

The over-limiting conductance has a weak dependence on the salt concentration. In 1 M $CuSO_4$, both membranes show almost the same limiting current without any OLC (Figure S1A), as expected for classical electro-diffusion. This is consistent with the SC mechanism since the over-limiting conductance is proportional to surface charge, which decreases at high salt[35], and the ratio of surface to bulk conduction scales with the inverse salt concentration[45]. On the other hand, in dilute 0.1 mM $CuSO_4$, AAO(-) shows a higher current than the AAO(+), although the current decreases as the potential is increased due to the extremely low concentration of $Cu^{2+}$ cations (Figure S1B). Comparing currents at the same voltage, the relative OLC for AAO(-) decreases weakly with salt concentration (Figure S1C), as expected theoretically for the SC mechanism. In contrast, both theory[45] and experiments[35] show that the over-limiting conductance increases significantly with salt concentration for the EOF mechanism.

The variation of potential with time at constant applied currents in 10 mM $CuSO_4$ also demonstrates the importance of SC in nanochannels (Figure S2). Below the limiting current (-0.5 mA and -1 mA), the potential variation is almost the same regardless of surface charge (Figure S2C), again confirming the dominance of bulk electrodiffusion over SC. When the applied current is close to the limiting current (-1.5 mA and -2 mA), AAO(+) shows an abrupt potential increase within ~100s (Figure S2A). The higher the applied OLC (-3 mA and -4 mA) is, the shorter the time at which the rapid increase in the potential occurs. This supports the interpretation that OLC in the AAO(+) generates an ion depletion region in front of the cathode, leading to a large overpotential that can cause side reactions, such as water electrolysis, consistent with observed gas bubbles. In contrast, AAO(-) maintains a low potential around -100 mV under -4 mA (Figure S2B), which shows that SC can sustain the electrodeposition process during OLC.

The dominant transport processes are also confirmed by impedance spectroscopy (Figure 3). Different direct currents are applied together with an alternating current of amplitude of 10 µA in the frequency range 100 kHz to 0.1 Hz. Figure 3A shows the Nyquist plots for varying surface charge and applied current. (The full-scale Nyquist plot and Bode plots are shown in Figure S3.) When -0.5 mA is applied, the impedance is almost independent of the surface charge, except that the total Warburg-like resistance of AAO(+) is larger than that of AAO(-) by 6%, which is precisely the surface-to-volume ratio of the pore, estimated as the area fraction of the EDL, $\lambda_D / h_p = 0.06$, where $\lambda_D = 5.0$ nm is the Debye length. This supports our hypothesis that the surface charge dependence results from SC asymmetry for the active $Cu^{2+}$ ions, even below the limiting current. Under -1 mA, the Warburg-like arc for both cases shrinks, consistent with a shortening of the diffusion layer, as the depleted zone expands into the pore.

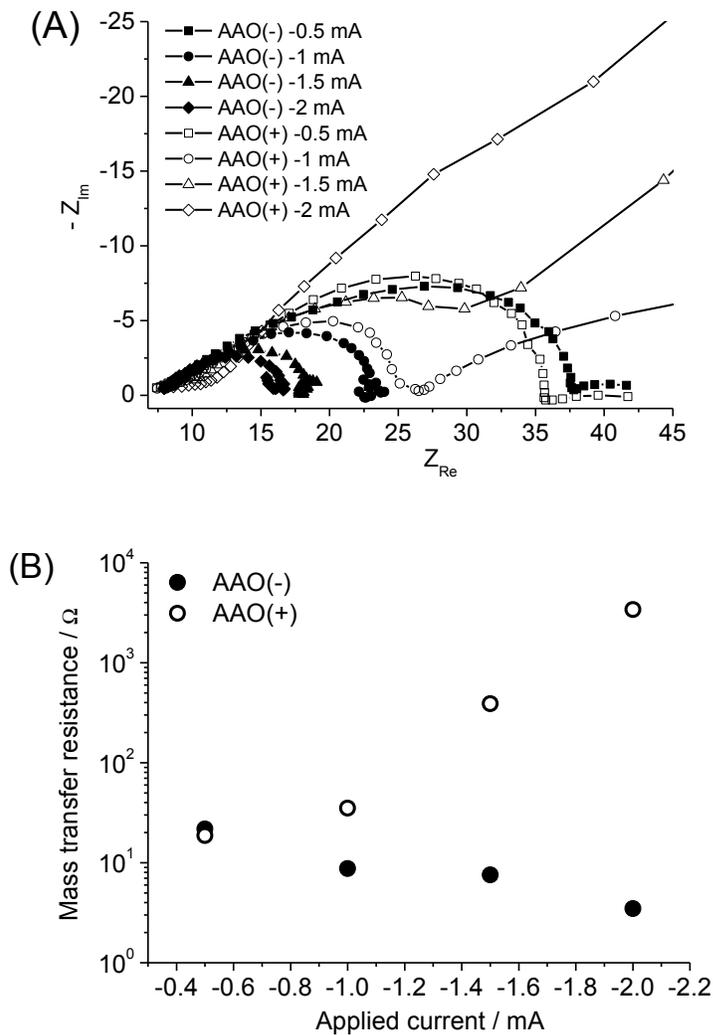

**Figure 3.** (A) Nyquist plots of AAO (+) and AAO (-) with different direct currents in 10 mM $CuSO_4$, (B) Fitted mass transfer resistance versus current. The resistance of AAO (+) at -1.0 mA includes both $R_{bd}$ and $R_{sc}$.

The impedance at high currents further supports our physical picture (Figure 2B). For the AAO(-), there is no other impedance feature, consistent with a negligible resistance for SC in the depleted region, and the Warburg-like arc shrinks with increasing current. For AAO(+), a new low-frequency feature develops for -1 mA that overwhelms the diffusion arc below -1.5 mA and leads to orders-of-magnitude larger mass-transfer resistance versus AAO(-). (See Figures 3 and S3A.) This indicates significant ion blocking by SC in AAO(+), also confirmed by imaging the electrodeposit below.

Our interpretation of the impedance spectra is quantified by fitting to four equivalent circuit models (Figure S4), depending on the applied current and surface charge of AAO. These models consist of the solution resistance ($R_s$), charge transfer resistance ($R_{ct}$), bulk diffusion resistance ($R_{bd}$), constant phase element (CPE), and additional resistance ($R_{sc}$) and pure capacitance (C) due to SC. The CPE is introduced to take into account the surface roughness of the electrode and/or the inhomogeneous reaction rate. (The fitted Nyquist plots are shown in Figure S5.) These models are necessarily empirical since there is no theory available for electro-diffusion impedance in a charged nanopore during OLC (unlike the case below limiting current[70]), but they suffice to extract consistent trends, such as the total mass transfer resistance versus the applied current (Figure 3B). AAO(-) maintains low resistance due to SC-driven OLC that decreases with increasing current, which we attribute to the shrinking diffusion layer as the depletion zone expands. On the other hand, the resistance of AAO(+) diverges as the current is increased, indicating severe ion depletion.

Our physical picture (Figure 3B) is further supported by the morphology of copper deposits grown during OLC, which reveals for the first time the dramatic effects of nano-template surface charge (Figure 4A). In the SC-dominated regime, we expect AAO(+) to block copper penetration into the nanopores, while AAO(-) should promote growth of a nanowire array following a deionization shock that is stable to shape perturbations[55]. For sufficiently high voltage and low salt, SC-guided electrodeposition should conformally coat the surfaces, leading to an array of nanotubes.

In order to test these theoretical predictions, copper electrodeposits are grown under OLC of -6 mA, three times the limiting current (-2 mA). In these experiments, the cathode is copper evaporated on a silicon wafer in order to facilitate subsequent cross-sectional scanning electron microscopy (SEM). As soon as the current is applied, both AAO(+) and AAO(-) show a drastic increase of potential after 20 s (Figure 4B-C), influenced by the kinetics of Cu reduction and nucleation. The potential for AAO(+) is unstable and reaches a much larger value, -1.75 V, leading to gas bubbling, while AAO(-) exhibits a stable, low potential around -0.1 V.

The morphology of the deposits is revealed by SEM images (Figure 4D-E), and their composition is confirmed to be pure copper by energy dispersive x-ray spectroscopy (EDS) (Figure S6). In AAO(-), an array of nanowires is obtained with an average length of 35 μm, set by the time of the experiment. In stark contrast, the growth in AAO(+) extends less than 3 μm into the nanopores (< 10 times their diameter), during the same experimental time. The positive surface charge effectively blocks dendritic growth from entering the porous template, leading to uniform copper electroplating below the template (not shown). Consistent with the theory[35, 45], this striking effect of surface charge is reduced by increasing salt concentration. In 1 M $CuSO_4$, the copper nanowires in the negative AAO are only slightly longer than that in the positive AAO because the SC is less important compared to bulk electrodiffusion in a concentrated electrolyte (Figure S7). These results show, for the first time, that electrodeposition in nanopores can be controlled by varying the surface charge, salt concentration, and current to change the relative importance of bulk and surface transport.

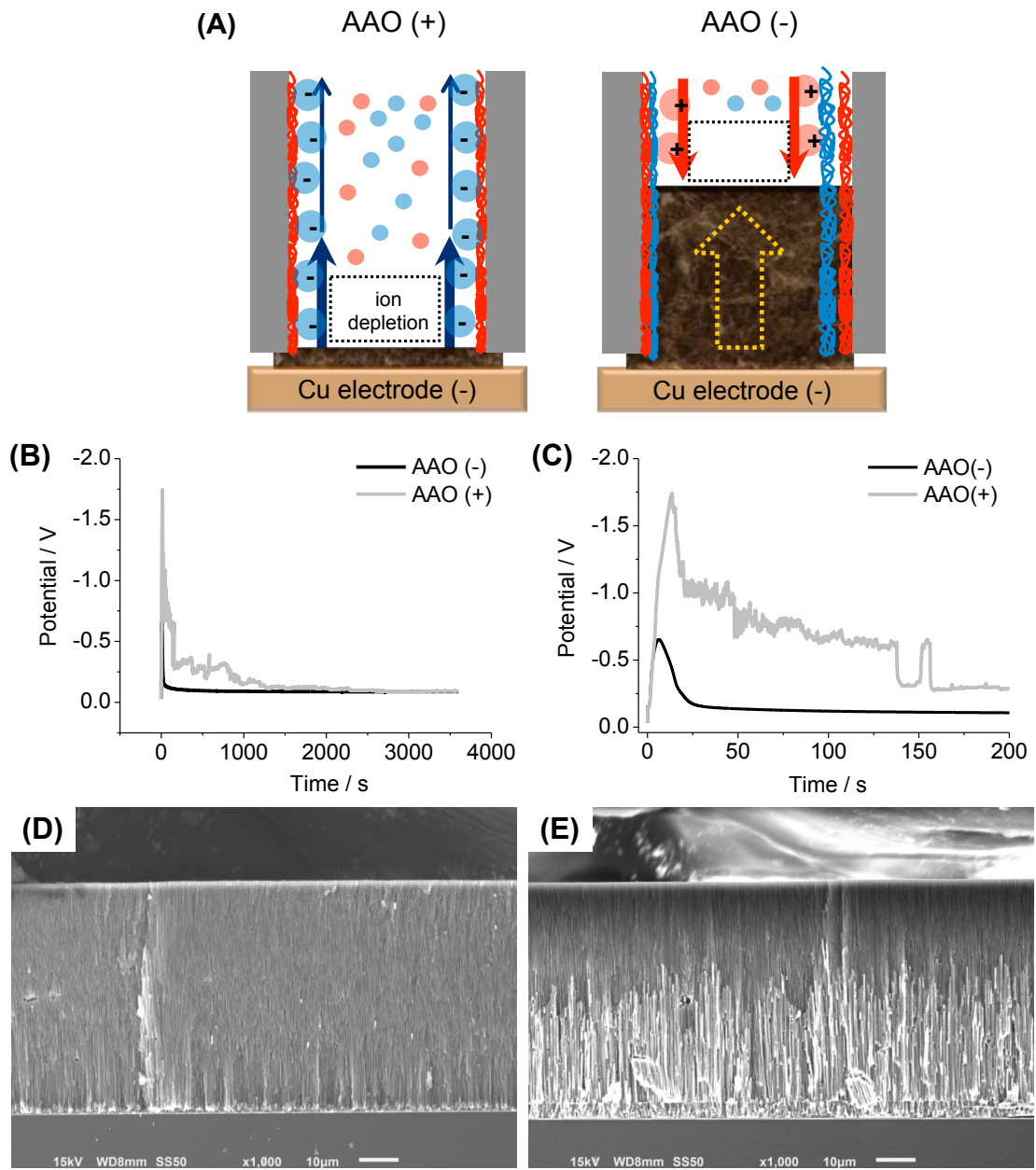

**Figure 4.** (A) Effect of SC on electrodeposition in charged nanopores during OLC. (B) V-t curves of AAO(+) and AAO(-) for an applied current of -6 mA. (C) Magnification of data of (B) for first 200 s. SEM images of electrodeposited Cu nanowires in (D) AAO(+) and (E) AAO (-).

Nanotubes grown over the surface of AAO(-) provide visual evidence of the SC mechanism. Although we find some nanotubes in the original experiments, more consistent nanotubes are obtained at higher voltages (further into the SC dominated regime) by chronoamperometric electrodeposition in a three-electrode cell, where AAO/Cu-evaporated on a Si wafer is used as the working electrode. A graphite pole and Ag/AgCl electrode are used as counter and reference electrodes, respectively, in order to accommodate hydrogen evolution at the anode. $H_3BO_3$ is added to reduce the hydrogen evolution rate at a high voltage and does not affect SC-driven OLC (Figure S8). To attach the AAO template to the Cu-evaporated Si wafer electrode, pre-electrodeposition is carried out in a two-electrode cell (Figure 1A) in 100 mM $CuSO_4$/100 mM $H_3BO_3$ by employing repeating chronopotentiometry, where underlimiting current (-10 mA) and 0 mA are applied for 30 s and 15 s respectively for 20 cycles. SEM images confirm that the height and the morphology of pre-electrodeposits are almost the same regardless of surface charge of AAO membranes. After pre-electrodeposition, the three-electrode cell is arranged and a large voltage, -1.8 V, is applied in the same electrolytic solution.

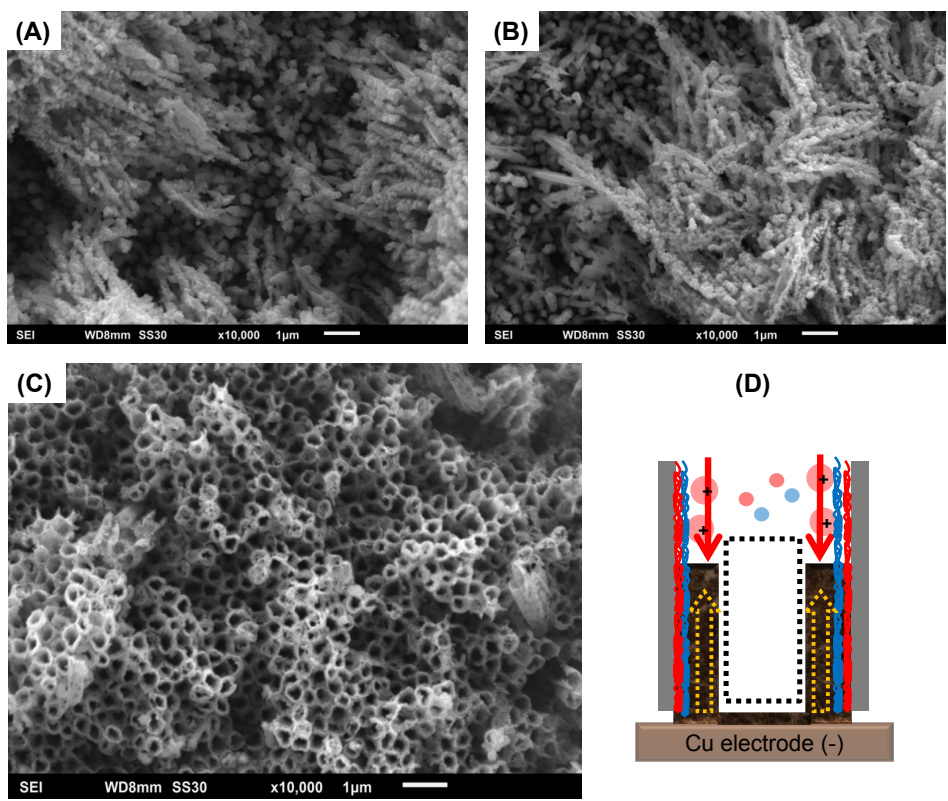

**Figure 5**. The effect of SC on the morphology of copper electrodeposits grown in 100 mM $CuSO_4$/100 mM $H_3BO_3$ solution after -1.8 V is applied for 5 min. SEM images of irregular nanowires generated in (A) bare AAO and (B) AAO(+). (C) SEM image of nanotubes grown in AAO(-), driven by SC as in (D).

Figure 5 shows the dependence of the electrodeposit morphology on the nanopore surface charge, far above the limiting current. The bare AAO and AAO(+) have irregular nanowires (Figure 5A-B). Note that the surface of bare AAO is slightly positive since the isoelectric point (pI) of aluminum oxide is around 8. The irregular dendritic growth, penetrating past the blockage demonstrated in Figure 4D, may result from electroconvection in the depleted region at this high voltage. On the other hand, AAO(-) at the same voltage shows well-defined copper nanotubes of uniform height (Figure 5C and Figure S9), whose wall thickness is less than 20 nm (Figure S10). This is consistent with SC control (Figure 5D) rather than previously proposed mechanisms that are independent of the surface charge, such as chemical affinity[71], vertical current by high current or potential[21], and morphology of sputtered metal[20].

Figure 6 illustrates the high-voltage morphological transitions. At -1.0 V, rough nanowire growth is observed that penetrates in four minutes less than 2 µm for AAO(+), compared to 3 µm in AAO(-). At -1.3 V, surface dendrites fed by SC growing along the pore walls to 4-5 µm in AAO(-), while longer, thin dendrites grow to 5-6 µm in AAO(+), avoiding the walls due to opposing SC. At -1.5 V, the surface dendrites in AAO(-) become more dense and transition to conformal-coating nanotubes reaching 6-7 µm, while those in AAO(+) are guided along the pore center out to 5-8 µm without touching the walls. In contrast to random, fractal growth in bulk solutions, dendrites can be precisely controlled in nanopores by tuning the surface charge, voltage and geometry.

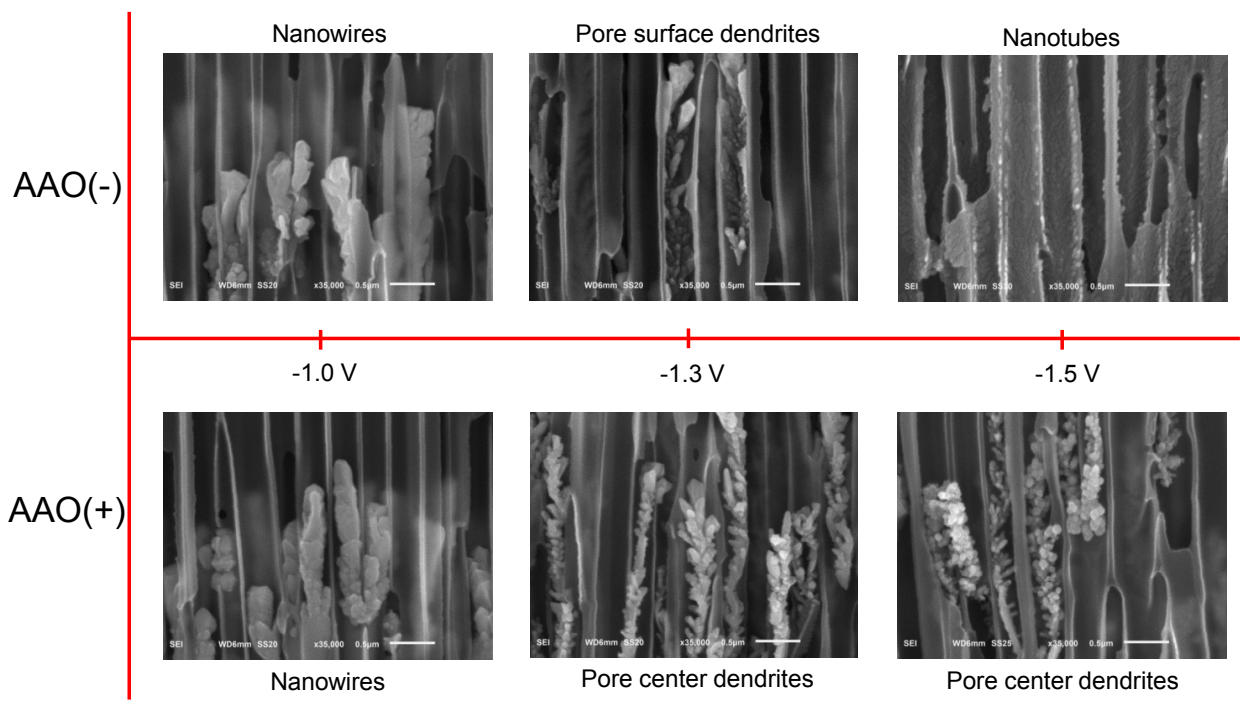

**Figure 6.** High-resolution SEM images (with 0.5 µm scale bars) of copper electrodeposits grown in charged AAO membranes, showing the morphology transition versus pore surface charge and the applied voltage. Electrodeposition was carried out in 100 mM $CuSO_4$/100 mM $H_3BO_3$ at each potential for 4 min.

In summary, this appears to be the first experiment demonstrating the importance of surface transport in electrodeposition. By modulating the surface charge of AAO nanopores with polyelectrolytes, we show that surface conduction (SC) is responsible for either enhancement or suppression of over-limiting current (OLC) between copper electrodes, depending on the sign of the surface charge. For positive surface charge (same as the electro-active copper ions), SC blocks dendrite penetration upon ion depletion; at high voltage, dendrites are channeled along the pore centers, avoiding the double layers. For negative surface charge, SC promotes uniform electrodeposition into the AAO template during OLC; at high voltage, growth is guided along the pore walls, consistent with an observed transition from copper nanowires to nanotubes.

These findings have many possible applications in electrochemical systems, microelectronics, and nanotechnology. SC-guided electrodeposition in nanopores could be used in place of solid electrolyte breakdown for programmable-metallization[16] or resistive-switching[17] random access memory, a low-voltage alternative to flash memory where each bit is a copper dendrite that reversibly short circuits two nanoelectrodes and acts like a memristor[18, 19]. Surface charge modification can also be used to control the morphology of metal electrodeposition in nanostructured templates for 3D electronics[17, 23, 26, 27], 3D batteries[28-30], and nanostructure synthesis[20, 21, 23, 25]. By selectively coating polyelectrolytes or other charged species on a template by lithography[23], patterns of suppressed or enhanced electrodeposition with desired morphology can be achieved. By dissolving the template after growth, multifunctional nanoparticles for electrocatalysis[24], molecular sensing or material additives can be made by combining metals, nanoparticles, polymers, and polyelectrolytes during SC-guided electrodeposition. The possibility of suppressing metal growth with positively charged coatings in porous media could also have applications to dendrite-resistant battery separators and reversible metal anodes for rechargeable batteries[6,8], in contrast to the negatively charged separators considered in recent work[12-14, 54]. Finally, this work highlights the need for new models of electrodeposition in porous media that account for electric double layers.

**Methods Summary**

Materials: All chemicals including poly(allylamine hydrochloride) (PAH, 15000 $M_w$), poly(styrenesulfonate) (PSS, 70000 $M_w$), copper sulfate ($CuSO_4$), sodium chloride (NaCl), hydrochloric acid (HCl), boric acid ($H_3BO_3$) and sodium hydroxide (NaOH) were purchased from Sigma-Aldrich and used without further purification. Ultrapure deionized water was obtained from Thermo Scientific (Model No. 50129872 (3 UV), Thermo Scientific). AAO membranes (pore diameter 300-400 nm, thickness 60 μm, length 47 mm, porosity 0.25-0.50) were purchased from Whatman (No. 6809-5022).

Electrode preparation: Two copper (Cu) disk electrodes (diameter 13 mm, thickness 2 mm) were used as the working and counter electrodes. Electrode polishing consisted of grinding by fine sand paper (1200, Norton) followed by 3.0 μm alumina slurry (No. 50361-05, Type DX, Electron Microscopy Sciences) and thorough rinsing with purified water.

Instruments: All electrochemical measurements were performed with a potentiostat (Reference 3000, Gamry Instruments). A pH meter (Orion 910003, Thermo Scientific) was used to adjust the pH of the polyelectrolyte solution. The morphology and composition of electrodeposited Cu nanostructures were confirmed by scanning electron microscopy (SEM) with energy-dispersive X-ray spectroscopy (EDS) detector (6010LA, JEOL) at 15 kV accelerating voltage.

Layer-by-layer deposition within AAO membrane: The AAO membrane was treated under air plasma for 5 min to generate a negative charge. The negatively charged AAO was immersed in a polycationic solution (1 mg/mL PAH in 20 mM NaCl at pH 4.3) for 30 min to generate a positive surface charge. Next, the membrane was thoroughly rinsed with purified water three times (10 min for each rinse) to remove unattached polyelectrolytes. The PAH-coated AAO was immersed in a polyanionic solution (1 mg/mL PSS in 20 mM NaCl at pH 4.3), followed by the same cleaning step. The polyelectrolyte-coated AAO was stored in $CuSO_4$ solution. The AAO template was dissolved with 1 M NaOH solution for 2 hours to get front images of Cu dendrites.

**Acknowledgements**

J.-H. H. acknowledges support from the Basic Science Research Program of the National Research Foundation of Korea funded by the Ministry of Education (2012R1A6A3A03039224), E. K. from a National Science Scholarship (PhD) funded by Agency of Science, Technology and Research, Singapore (A*STAR), and M. Z. B. from an IBM Faculty Award. Additional support came from Saint Gobain Ceramics and Plastics, Northboro Research and Development Center. The authors thank Sunhwa Lee and Prof. Paula Hammond for advice about the layer-by-layer method and Dr. Ramachandran Muralidhar for discussions on CB-RAM.

**Author Contributions**

All authors contributed to the research. J.-H. Han led the experiments and analyzed the data. E. K. derived the equations and performed the numerical simulations. P.B. helped design the electrochemical cell. M. Z. B suggested the approach and led the theoretical interpretation and writing.

**Additional information**
Data fitting, V-t curves, Bode plots, equivalent circuit models, Nyquist plots, and SEM images are included in the supporting information.

**Competing financial interests**
The authors declare no competing financial interests.

# Next page: additional figures.

# Supporting Information

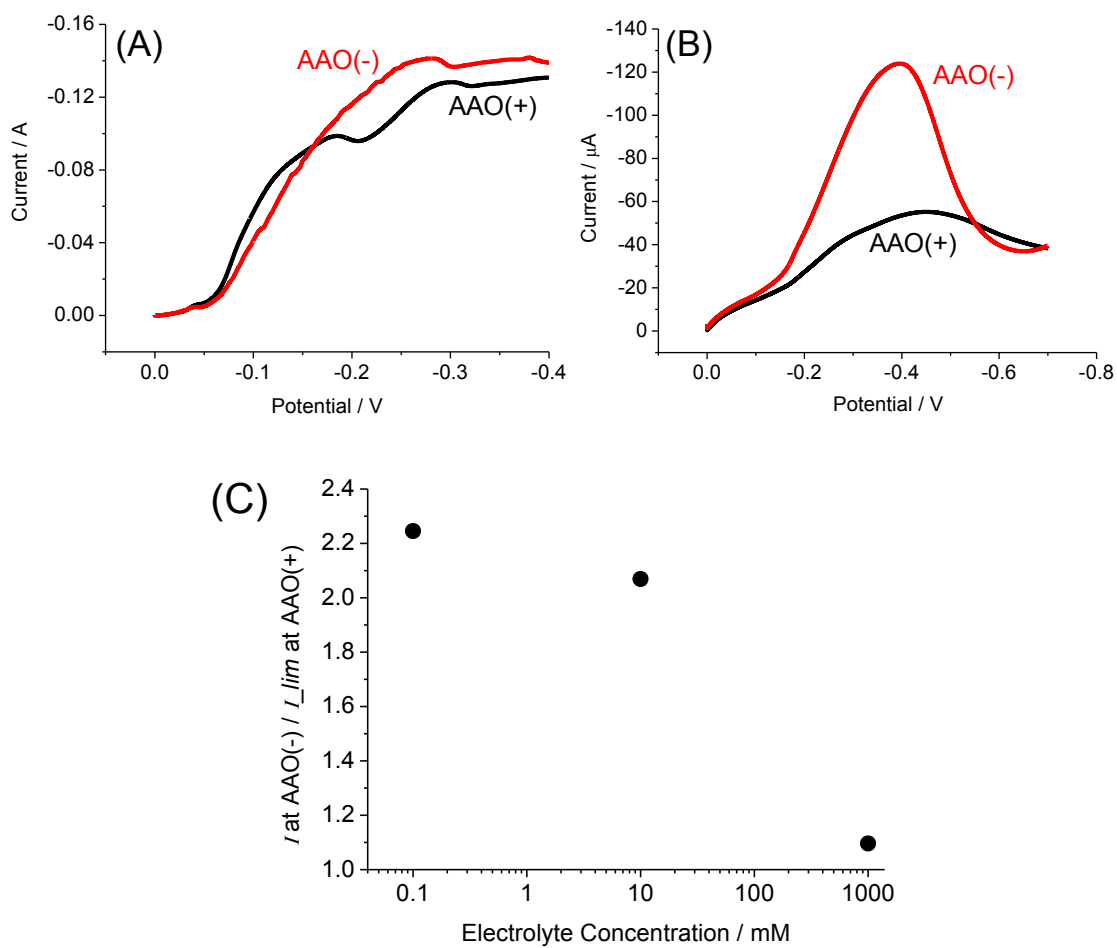

**Figure S1.** The effect of electrolyte concentration on SC-driven OLC. I-V curves of AAO(+) and AAO(-) membranes in (A) 1 M $CuSO_4$ and (B) 0.1 mM $CuSO_4$ at a scan rate of 1 mV/s. (C) A plot of current ratio of AAO(-) to AAO(+) as a function of electrolyte concentration.

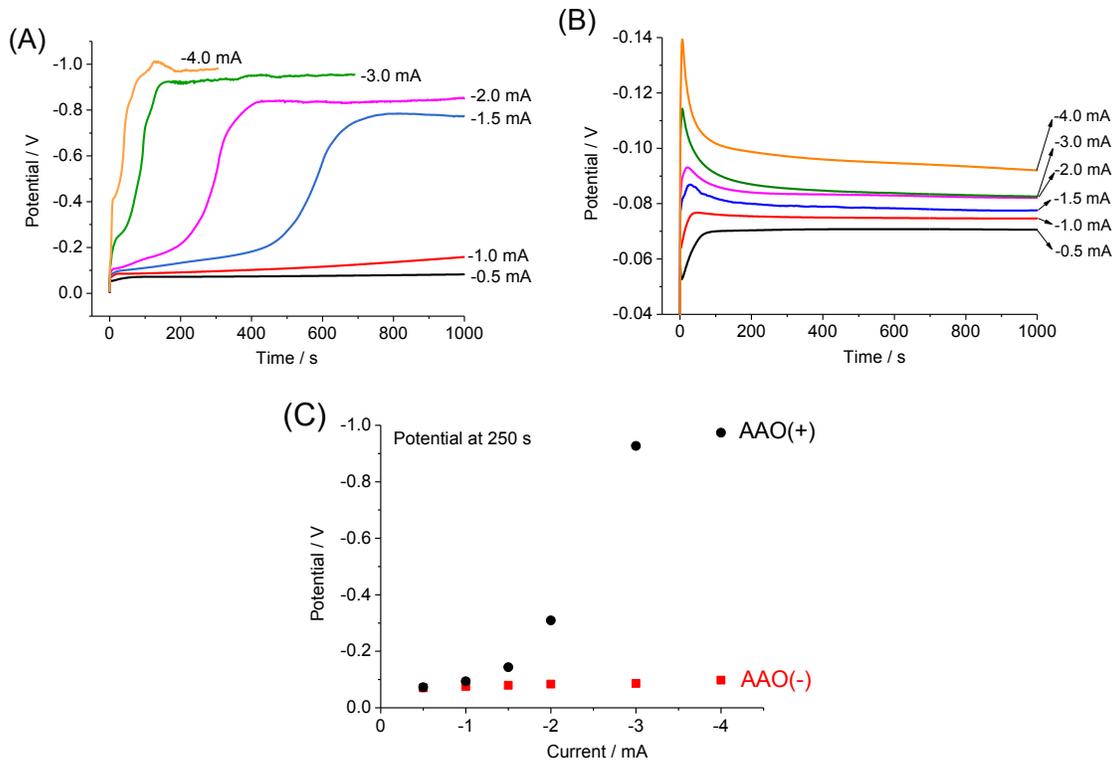

**Figure S2.** V-t curves of (A) AAO(+) and (B) AAO(-) in 10 mM CuSO$_4$ with different applied currents. (C) A comparison of potential at 250 s as a function of surface charge and applied currents.

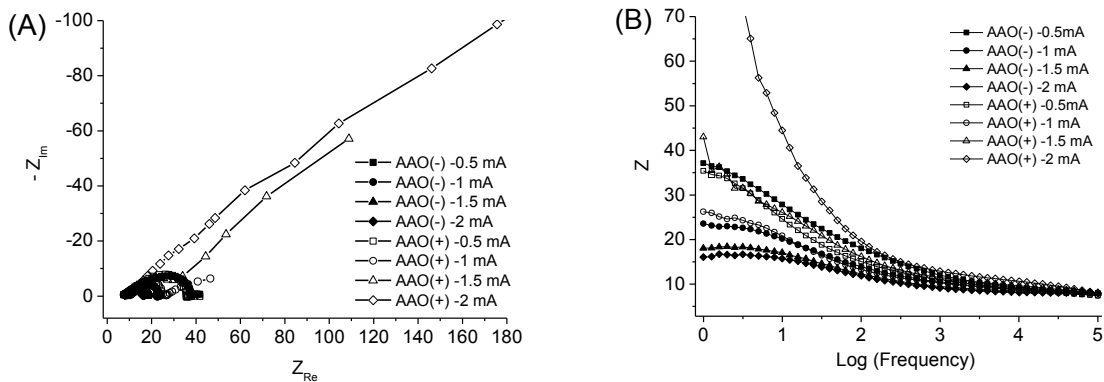

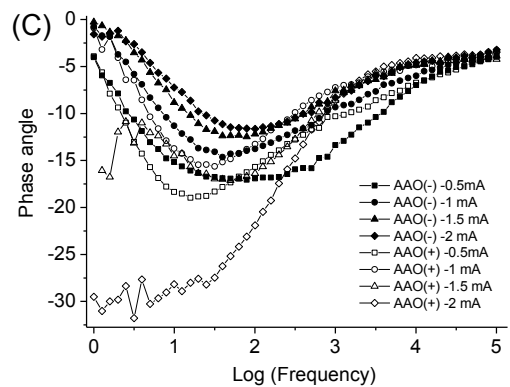

**Figure S3.** Full-scale Nyquist plots (A) and Bode plots (B and C) of AAO(+) and AAO(-) with different direct currents in 10 mM $CuSO_4$. (B) Bode plots of total impedance, and (C) Bode plots of phase angle.

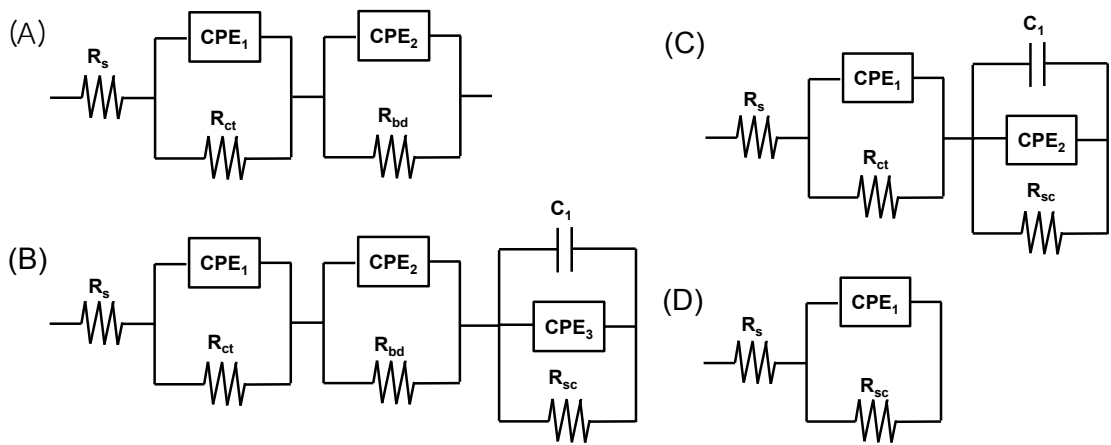

**Figure S4.** Four kinds of equivalent circuit models for (A) AAO(-) at all currents and AAO(+) at -0.5 mA, (B) AAO(+) at -1.0 mA, (C) AAO(+) at -1.5 mA and (D) AAO(+) at -2 mA.

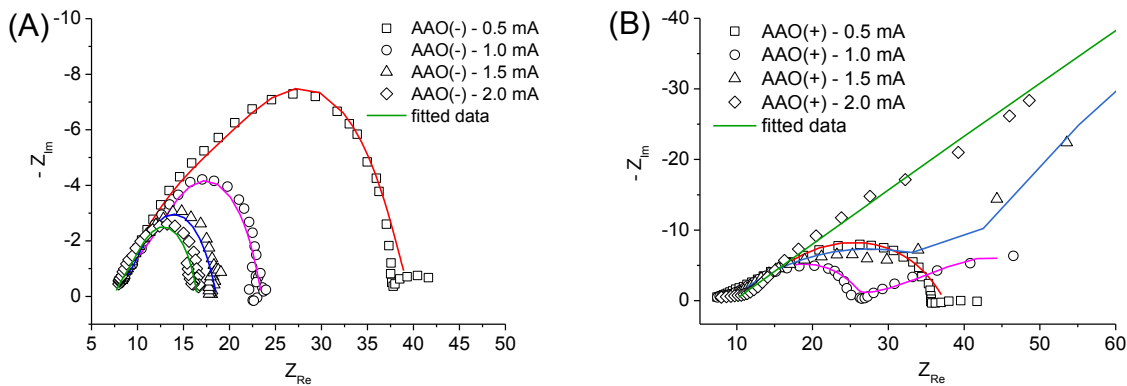

**Figure S5.** Nyquist plots of (A) AAO(-) and (B) AAO(+) with fitted data. The dotted and solid lines are experimental data and fitted data, respectively.

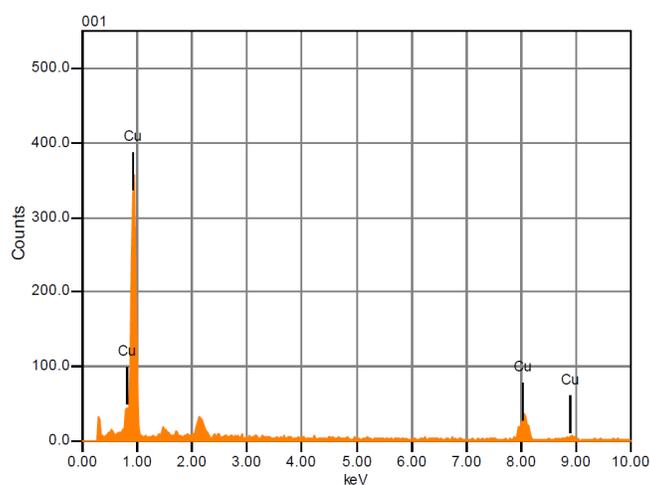

**Figure S6.** EDS data of Cu nanowire arrays

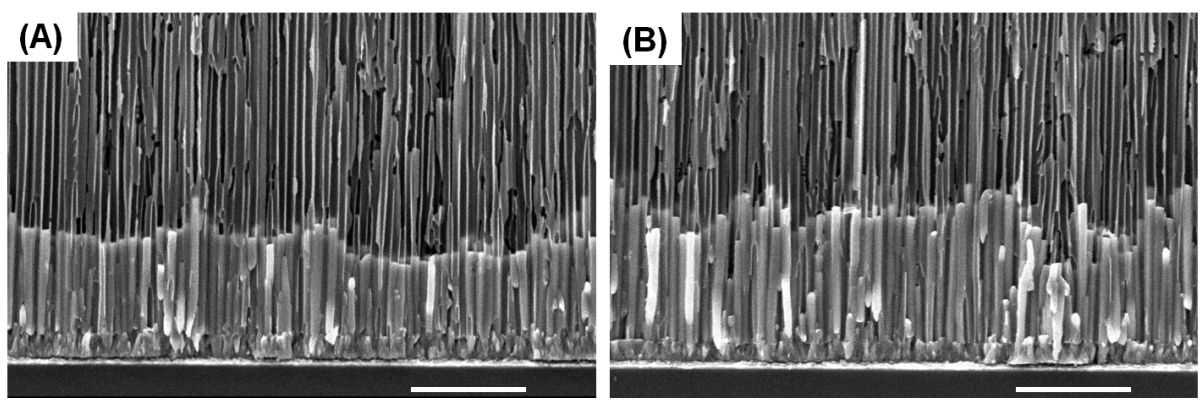

**Figure S7.** SEM images of Cu nanowires electrodeposited from (A) AAO(+) and (B) AAO(-) in 1 M CuSO$_4$. The scale bars are 5 μm.

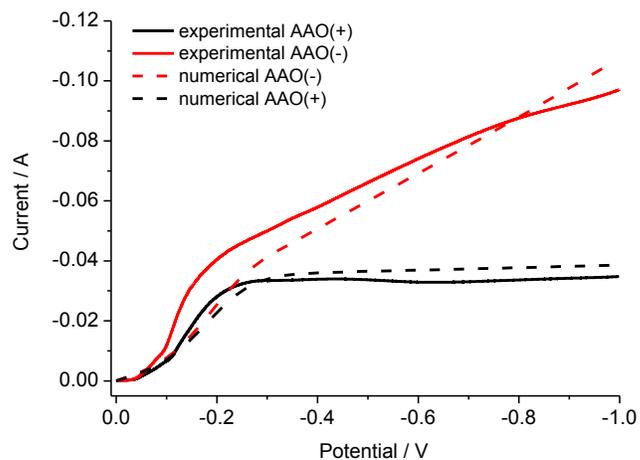

**Figure S8.** Current data for AAO(+) and AAO(-) membrane in 100 mM CuSO$_4$/100 mM H$_3$BO$_3$ at a scan rate of 10 mV/s.

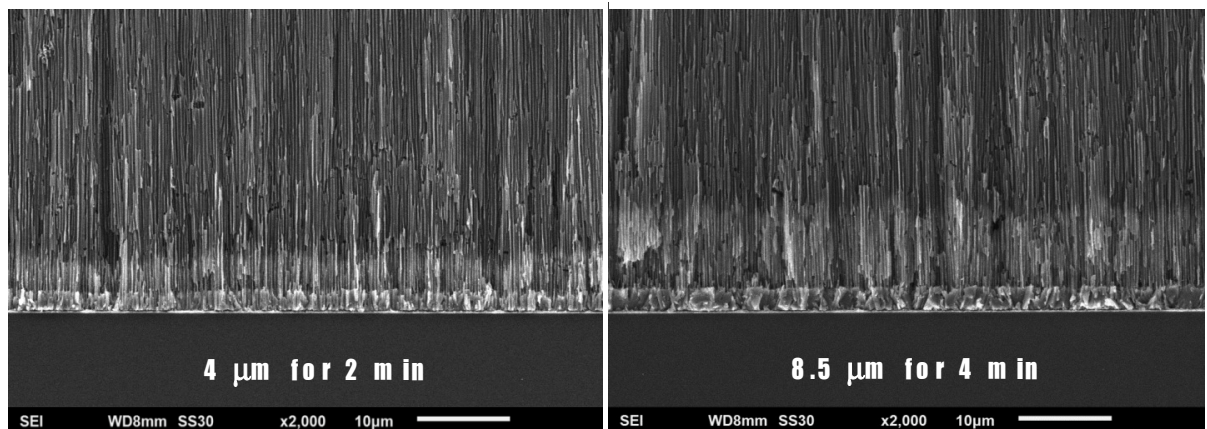

**Figure S9.** Cross-sectional SEM images of nanotubes generated in AAO(-) membrane. -1.8 V was applied for 2 and 4 min in 100 mM $CuSO_4$ /100 mM $H_3BO_3$ solution at room temperature. The nanotubes are very uniform and growth rate is about 2 µm per min.

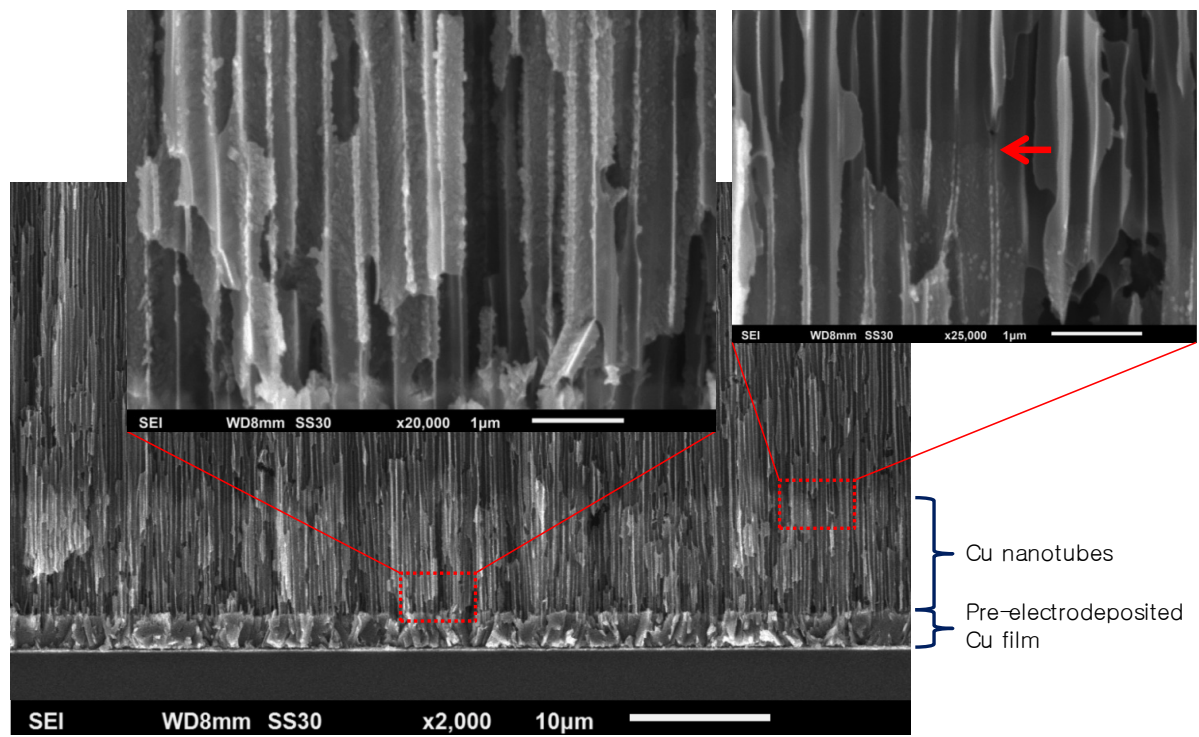

**Figure S10.** Cross-sectional SEM images of nanotubes generated in AAO(-) membrane. -1.8 V was applied for 4 min in 100 mM $CuSO_4$ /100 mM $H_3BO_3$ solution at room temperature. The red arrow indicates the tip of nanotubes along the walls of AAO membrane.